\documentclass{article}

\usepackage{amssymb,amsfonts,amsmath}
\usepackage{cite,enumerate,float,indentfirst}
\usepackage{color}

\def\be{\begin{eqnarray}}
\def\ee{\end{eqnarray}}
\def\nn{\nonumber}

\def\p{\partial}

\def\Tr{{\rm Tr}\,}

\def\l[{\phantom.[}

\def\ttau{Z}

\hoffset= - 1.0in         

\begin{document}

\title{\vspace{-.1cm}{\Large {\bf On determinant representation and integrability
of Nekrasov functions}\vspace{.2cm}}
\author{
{\bf A.Mironov$^{a,b,c}$}\footnote{mironov@lpi.ru; mironov@itep.ru}\ \ and
\ {\bf A.Morozov$^{b,c}$}\thanks{morozov@itep.ru}}
\date{ }
}

\maketitle

\vspace{-5.5cm}

\begin{center}
\hfill FIAN/TD-13/17\\
\hfill IITP/TH-12/17\\
\hfill ITEP/TH-20/17
\end{center}

\vspace{3.3cm}

\begin{center}
$^a$ {\small {\it Lebedev Physics Institute, Moscow 119991, Russia}}\\
$^b$ {\small {\it ITEP, Moscow 117218, Russia}}\\
$^c$ {\small {\it Institute for Information Transmission Problems, Moscow 127994, Russia}}

\end{center}

\vspace{.5cm}

\begin{abstract}
Conformal blocks
and their AGT relations to LMNS integrals and Nekrasov functions
are best described by "conformal"
(or Dotsenko-Fateev) matrix models, but in non-Gaussian
Dijkgraaf-Vafa phases, where different eigenvalues are
integrated along different contours.
In {\it such} matrix models, the determinant representations
and integrability are restored
only after a peculiar Fourier transform in the numbers of
integrations.
From the point of view of conformal blocks, this
is Fourier transform w.r.t. the intermediate dimensions,
and this explains why such quantities are expressed
through tau-functions in Miwa parametrization,
with external dimensions playing the role of  multiplicities. In particular, these determinant representations provide solutions to the Painlev\'e VI equation.
We also explain how this pattern looks
in the pure gauge limit,
which is described by the Brezin-Gross-Witten matrix model.
\end{abstract}

\bigskip

\bigskip

\section{Introduction}

AGT relations \cite{AGT}, identifying the conformal blocks and the Nekrasov functions,
possess different interpretations.
The most straightforward and useful one is through properly defined
Dotsenko-Fateev-like (DF) integral representations of the conformal blocks \cite{DF,confmamo,AGTmamo,MMSha},
which can be interpreted as matrix models,
with character decompositions \cite{charMM,MMS} looking exactly
like the Nekrasov sums over representations.
Matrix models possess a lot of other nice properties,
which can be then transmitted to either the Nekrasov functions or to the conformal blocks,
especially at $c=1$ (i.e. $\beta=1$).
Among these features are the closely related integrability and
determinant representations, see \cite{UFN3} for comprehensive reviews and
references.
Particular pieces of this general pattern are constantly being rediscovered
in particular studies of particular questions.

Let us note that the DF representation of the conformal block leads to quite a complicated matrix model: a $\beta$-ensemble in the non-Gaussian Dijkgraaf-Vafa (DV) phase \cite{DV}. This means that the different eigenvalues are integrated along different contours. In this paper, we restrict ourselves with the case of unit central charge, $c=1$ in the conformal theory, which means that we deal not
with a  $\beta$-ensemble, but with an ordinary matrix model.
In fact, a lot of general properties discussed below  persist for the $\beta$-ensembles too.

In the Dijkgraaf-Vafa phases of matrix models, the determinant representations and integrability are restored only after a peculiar Fourier transform in the numbers of integrations \cite{MMZ}. From the point of view of conformal blocks, this is Fourier transform w.r.t.
(the square roots of) the intermediate dimensions, and this explains why such quantities are expressed through tau-functions in Miwa parametrization, with external dimensions playing the role of  multiplicities.
Moreover, these determinant representations provide solutions to the Painlev\'e VI equation.

Strictly speaking,
the matrix model representations exist only when two integrality conditions are imposed on the conformal momenta
\be\label{cn2}
N_1=\alpha-\alpha_1-\alpha_2,\ \ \ \ \ N_2=-\alpha-\alpha_3-\alpha_4
\ee
while the conformal block at generic values of the external dimensions is obtained by the analytic continuation.
This analytic continuation is immediate for various expansions of the conformal block \cite{MMM}, but not that immediate for determinant representation,
since it implies that determinant can be of a matrix of non-integer size.
One possibility to handle this situation is to change a matrix determinant
for an infinite-dimensional operator determinant.

This idea was realized on the other side of the AGT story, where there is a long program \cite{GIL1}-\cite{GL} of
interpreting linear combinations of conventional conformal blocks
in terms of Painlev\'e $\tau$-functions.
Two facts were revealed in these papers:
that a Fourier transform of the conformal block in the intermediate conformal momentum admits a Fredholm determinant representation, and that it satisfies the Painlev\'e VI equation.
These claims were actually made only for the case, when the conformal momenta satisfy
\be\label{cn1}
\alpha_1\pm\alpha_2+\alpha\notin\mathbb{Z},\ \ \ \ \ \ \ \alpha_1\pm\alpha_2-\alpha\notin\mathbb{Z},\ \ \ \ \ \ \ \alpha_3\pm\alpha_4+\alpha\notin\mathbb{Z},\ \ \ \ \ \ \  \alpha_3\pm\alpha_4-\alpha\notin\mathbb{Z}
\ee
which is a kind of complementary   to (\ref{cn2}) in the matrix model approach.
In fact these complicated functional determinants are nothing more than generalizations
of the {\it finite} ones, made from very simple hypergeometric functions,
which arise at the "integer" locus (\ref{cn2}).

It turns out that these properties persist \cite{GIL2,ILT3,BS1,GL} in the "pure-gauge" limit
of AGT relations, which is somewhat peculiar in many respects.
Most important, the relevant matrix model is the celebrated Brezin-Gross-Witten (BGW) model
\cite{BGW}, which was studied in great detail in \cite{MMSem,AMM},
where it was shown to possess determinant representation in terms of the
Bessel functions, see also \cite{Bessel}. In fact, there is an even more interesting matrix model representation of the pure gauge limit, \cite{Grassi} which is, however, different from the framework described in this paper, and deserves a separate discussion.

In the pure gauge limit (PGL), the ${\cal N}=2$ SUSY Yang-Mills theory is no longer
conformal, due to dimensional transmutation one trades masses
of the hypermultiplets for a new parameter $\Lambda$.
From the point of view of conformal theory, this corresponds to pushing all
external dimensions to infinity, while simultaneously approaching the
singularity of the conformal block: this eliminates both external dimensions
and puncture positions by a single $\Lambda$.
Moreover, according to \cite{ShaPGL}, in the PGL, the 4-point
spherical and 1-point toric conformal blocks coincide:
\be
B_*(\Delta|\Lambda) = \lim_{\stackrel{
\Delta_1,\Delta_2,\Delta_3,\Delta_4\longrightarrow \infty, \ q\longrightarrow 0}
{q(\Delta_2-\Delta_1)(\Delta_3-\Delta_4)\equiv \Lambda^4}}
B^{(0)}(\Delta_1,\Delta_2,\Delta_3,\Delta_4;\Delta,c|q)
= \lim_{\stackrel{\Delta_{ext}\longrightarrow \infty}{e^{i\pi\tau}\Delta_{ext}^2\equiv \Lambda^4}}
B^{(1)}(\Delta_{ext};\Delta,c|e^{i\pi\tau})
\ee

\bigskip

As we explained above, the conformal block at $c=1$ is described by the ordinary matrix model of
Penner type in Dijkgraaf-Vafa phase. We discuss the model in section 2. In section 3, we explain that appropriate Fourier transform of the conformal block provides the determinant representation, and in section 4, we discuss its integrability properties. In section 5, we explain that the Fourier transform satisfies the Painlev\'e VI equation in the both cases of conditions (\ref{cn2}) and (\ref{cn1}).

In the PGL, the Penner model is substituted by the BGW model, which is less investigated and requires a more detailed exposition. Hence, we remind the Shapovalov and character representations of Nekrasov functions in the PGL in ss.\ref{Shapo} and \ref{BGWMc} respectively. In sections 8 and 9, we discuss the integrability of the PGL, and explain that it satisfies the equation Painlev\'e III. Section 10 contains concluding remarks.

\section{Matrix-model description of conformal blocks \cite{MMSha}
\label{CBmamo}}

As explained in detail in \cite{MMSha}, by a suitable adjustment of Dotsenko-Fateev (DF)
trick \cite{DF} (applying it to holomorphic quantities and making use of open
rather than closed integration contours), conformal blocks can be converted
into the matrix-model form.
Emerging in this way are just the "conformal matrix models" of \cite{confmamo},
which are also close to Penner models \cite{Pen} and which are nowadays
naturally called DF-models.

As the simplest example, the 4-point conformal block (=Nekrasov function)  in the $c=1$ CFT \cite{CFT}
\be\label{CB}
B(\Delta_i;\Delta; q)=q^{\Delta-\Delta_1-\Delta_2 }\cdot\left( 1+ {(\Delta_2-\Delta_1+\Delta)(\Delta_3-\Delta_4+\Delta)\over 2\Delta}\cdot q+{\cal O}(q^2)\right)
\ee
can be realized via the matrix (eigenvalue) integral $Z_{N_1,N_2}$ \cite{AGTmamo}
\be
\boxed{
Z_{N_1,N_2}=\mathfrak{Z} \cdot B(\Delta_i;\Delta; q)}
\ee
\be\label{ZB}
Z_{N_1,N_2}=q^{2\alpha_1\alpha_2}(1-q)^{2\alpha_2\alpha_3}
\cdot{1\over N_1!N_2!}
\int\prod_i dx_i\Delta^2(x)\prod x_i^{2\alpha_1}(1-x_i)^{2\alpha_2}(q-x_i)^{2\alpha_3}
\ee
where the normalization factor
\be
\mathfrak{Z}=\prod_{i=1}^{N_1}{\Gamma (i)\Gamma \Big(2\alpha_1+i\Big) \Gamma \Big(2\alpha_2+i\Big)\over \Gamma \Big(2\alpha-i+1\Big)}\times
 \prod_{i=1}^{N_2}{\Gamma (i)\Gamma \Big(2\alpha_3+i\Big) \Gamma \Big(2\alpha_4+i\Big)\over \Gamma \Big(-2\alpha-i+1\Big)}
=C^\alpha_{\alpha_1\alpha_2}C^{-\alpha}_{\alpha_3\alpha_4}
\ee
with the structure constants
\be
C^\alpha_{\alpha_1\alpha_2}=
{\mathfrak{G}\Big(\alpha+\alpha_1-\alpha_2+1\Big)\mathfrak{G}\Big(\alpha+\alpha_2-\alpha_1+1\Big)\mathfrak{G}\Big(\alpha-\alpha_1-\alpha_2+1\Big)
\mathfrak{G}\Big(2\alpha\Big) \over\mathfrak{G}\Big(2\alpha_1+1\Big)
 \mathfrak{G}\Big(2\alpha_2+1\Big)\mathfrak{G}\Big(\alpha+\alpha_1+\alpha_2+2\Big)}
\ee
being nothing but the Selberg integrals \cite{MMSha,MMS} expressed through the Barnes $G$-functions $\mathfrak{G}(x)$ \cite{Barnes}, and the matrix integral (\ref{ZB}) depends on two integers, $N_1$ and $N_2$ that count the number of integrations over the contours $C_1=[0,q]$ and $C_2=[1,\infty)$ respectively:
\be
N_1=\alpha-\alpha_1-\alpha_2,\ \ \ \ \ N_2=-\alpha-\alpha_3-\alpha_4
\ee
and
\be
\Delta_i=\alpha_i^2,\ \ \ \ \ \ \ \Delta=\alpha^2
\ee
Thus,
\be
N=N_1+N_2=-\alpha_1-\alpha_2-\alpha_3-\alpha_4\equiv -\alpha_{1234}
\ee
parameterizes the fourth conformal dimension $\Delta_4$, while $N_1$,
the intermediate conformal dimension $\Delta$.
Generic values of $\alpha$'s correspond to non-integer $N_1$ and $N$,
but this analytical continuation is straightforward and unambiguous,
because (\ref{ZB}) belongs to the class of Selberg integrals,
which are ratios of polynomials and thus are well-defined analytical
functions of their variables \cite{MMSha,MMM,MMS} (see, however, \cite{IMMconf,minAGT} for description of more delicate situations).

However, at integer values of $N_1$ and $N_2$, additional structures emerge,
whose analytical continuation, though also straightforward is rather ugly.
These are determinant formulas underlying integrability properties.
We prefer to describe them in the "pure" case, at integer $N_{1,2}$,
when the determinants are finite-dimensional and $\tau$-functions look nice and simple.
The analytical continuation converts them into functional determinants,
for which there is still no nice terminology and commonly accepted condensed notation,
thus, one needs to write overloaded and non-transparent explicit formulas,
see \cite{GIL1}--\cite{GL} for examples.
In the next sections, we present the clear version of this story:
at integer values of $N_1$ and $N_2$.

\section{Determinant representation of Fourier-transformed matrix models \cite{MMZ}
\label{detrepDF}}

One can consider instead of $Z_{N_1,N_2}$ the standard $N$-fold matrix integral with all eigenvalues being integrated over the same contour that is a linear combination of the two contours $C_1$ and $C_2$,
\be\label{Zgf}
Z_N=q^{2\alpha_1\alpha_2}(1-q)^{2\alpha_2\alpha_3}
\cdot{1\over N!}
\int_C\prod_i dx_i\Delta^2(x)\prod_i x_i^{2\alpha_1}(1-x_i)^{2\alpha_2}(q-x_i)^{2\alpha_3}
\ee
with two generating parameters $\mu_1$ and $\mu_2$
\be
\int_C=\mu_1\int_{C_1}+\mu_2\int_{C_2}
\ee
This integral is clearly a  generation function of $Z_{N_1,N_2}$:
\be
\boxed{
Z_N(\mu_1,\mu_2)=\sum_{N_1,N_2:\ {N_1+N_2=N}}
\mu_1^{N_1}\mu_2^{N_2}\cdot Z_{N_1,N_2}
}
\label{preFT1}
\ee
since the binomial coefficient is cancelled by the normalization factorials in (\ref{ZB}) and (\ref{Zgf}).

For $Z_N(\mu_1,\mu_2)$ there is a determinant representation
\be
\boxed{
Z_N(\mu_1,\mu_2)= q^{2\alpha_1\alpha_2}(1-q)^{2\alpha_2\alpha_3}
\cdot\det_{1\le i,j\le N}G(i+j-2)
}
\label{detrep}
\ee
where
\be\label{G}
G(k)= \mu_1 \int_0^q x^{2\alpha_1+k}(1-x)^{2\alpha_2}(q-x)^{2\alpha_3}dx
+ \mu_2\int_1^\infty x^{2\alpha_1+k}(1-x)^{2\alpha_2}(q-x)^{2\alpha_3}dx =
\nn\\
=\mu_1\ q^{2\alpha_{12}+k+1}\ \mathfrak{B}(2\alpha_1+k+1,2\alpha_2+1)\phantom{A}_2F_1(-2\alpha_3,2\alpha_1+k+1;2\alpha_{12}+k+2;q)+\nn\\
+\mu_2\ \mathfrak{B}(-2\alpha_{123}-k-1,2\alpha_3+1)\phantom{A}_2F_1(-2\alpha_{123}-k-1,
-2\alpha_2;-2\alpha_{12}-k;q)
\label{Gfun}
\ee
where $\mathfrak{B}(\alpha,\beta)= \int_0^1 x^{\alpha-1}(1-x)^{\beta-1}dx=
{\Gamma (\alpha)\Gamma (\beta)\over\Gamma (\alpha+\beta)}$ is the standard Beta-function \cite{GR}, and the first and the second terms at the r.h.s. of (\ref{G}) are obtained by taking integrals of $x^{2\alpha_1}(1-x)^{2\alpha_2}(q-x)^{2\alpha_3}$ over $C_1$ and $C_2$ respectively.

In fact, the same trick with Fourier transform in the multiplicities $N_i$
of contour integrations is applicable to description of DV phases of generic
$\beta$-ensembles, i.e. to conformal blocks with $c\neq 1$.

\section{Toda chain equation in Miwa variables}

It is well known since \cite{GMMMO,versus} that the determinant (\ref{detrep}) is a
Toda chain $\tau$-function, see \cite{UFN3} for detailed explanations. More exactly, if one considers a generic matrix model
\be\label{Tc1}
Z_N={1\over N!}\int\prod_i df(x_i)\Delta^2(x)\prod_i e^{\sum_k t_kx_i^k}
\ee
with an arbitrary measure $f(x)$, then $Z_N$ is a Toda chain $\tau$-function with $N$ being the discrete Toda time variable, it has the determinant representation
\be
Z_N=\det_{1\le i,j\le N} C_{i+j-2},\ \ \ \ \ \ \ \ C_k\equiv \int df(x)\ x^k \exp\left(-{\mu x^2\over 2}+\sum_k t_kx^k\right)
\ee
and $Z_N$ satisfies the equations of the (forced) Toda chain hierarchy, the first of which is
\be
Z_N\partial^2 Z_N-\Big(\partial Z_N\Big)^2=Z_{N+1}Z_{N-1}
\ee
However, these equations are formulated in terms of the ordinary
time variables, which are not present in (\ref{detrep}).
Instead of infinitely many times, $G$ there depends just on three
$\alpha$-parameters, which are associated with the three points $\mu_1=0$, $\mu_2=q$, $\mu_3=1$ accordingly. They can be treated either as the measure $f(x)$ in (\ref{Tc1}), or as the Miwa variables. Let us choose the second option
and actually deal with the Toda chain $\tau$-function in terms of Miwa variables (this statement is immediately extended with more Miwa variables to multipoint conformal blocks). Hence, it satisfies the integrable equations (bilinear identities) in Miwa variables which we briefly remind here (see \cite{KMMMZ,versus} for details).

When converted from time to Miwa variables,
\be
t_k = \frac{1}{k}\sum_a p_a\mu_a^{-k}
\label{tMi}
\ee
the Hirota bilinear equations become 3-term difference equations with respect to the
multiplicities $p_a$ \cite{discHir}:
\be
(\mu_a-\mu_b)\, \tau[p_a,p_b,p_c+1]\,\tau[p_a+1,p_b+1,p_c] + \nn \\
+ (\mu_b-\mu_c)\,\tau[p_a+1,p_b,p_c]\,\tau[p_a,p_b+1,p_c+1] + \nn \\
+ (\mu_c-\mu_a)\,\tau[p_a,p_b+1,p_c]\,\tau[p_a+1,p_b,p_c+1] = 0
\label{HirMi}
\ee
and, at all unit multiplicities $p_a=1$, they are solved by
\be
\tau = Z_N= \frac{\det_{1\leq i,j\leq N} \phi_i(\mu_j)}{\Delta(\mu)}
\label{detrepMi}
\ee
with arbitrary set of functions of a single variables $\{\phi_i(\mu)\}$ with asymtotics at large $\mu$: $\phi_i(\mu) \sim \mu^{i-1}$.
Transition from (\ref{detrepMi}) to (\ref{HirMi})
involves taking a singular limit where several $p_a$ variables $\mu_a$ coincide.
As a byproduct of the study of this limit, one obtains another
interesting equation \cite{KMMMZ}:
\be
 p_a \ttau_{N+1}[p_a+1]\ \ttau_{N-1}[p_a-1]
 = \ttau_N^2 \frac{\p}{\p\mu_a} \frac{\hat\ttau_N[p_a]}{\ttau_N[p_a]}
\ee
where $\hat\tau_N$ differs from $\tau_N$ in (\ref{detrepMi})
by a substitution $\phi_N(\mu_j) \longrightarrow \phi_{N+1}(\mu_j)$
in the last row of the matrix at the r.h.s.

Coming back to the conformal blocks,
the function (\ref{Gfun}) would depend on time variables,
if there was a factor $\exp\left(\sum_{k=1}^\infty t_kx^k\right)$
in the integration measure.
Instead, the measure in (\ref{Gfun}) consists of three factors,
and parameters $\alpha_{1,2,3}$ are exactly the multiplicities $p_{1,2,3}$
in (\ref{tMi}) with
$\mu_1=0, \mu_2=q, \mu_3=1$.
Note that in (\ref{HirMi}) the $p$-variables do not need to be integer,
thus arbitrary complex-valued $\alpha_{1,2,3}$ are actually allowed.

Thus, we see that {\bf the Fourier-like transform of the conformal block at $c=1$
w.r.t. the square root $\alpha$ of the internal dimension
is the Toda chain $\tau$-function in Miwa parametrization,
with square roots of external dimensions playing the role of the
multiplicities}.
This may be considered a kind of underlying first principle of the observations of \cite{GIL1}--\cite{GL}.

\section{Painlev\'e VI equation for Fourier transformed conformal blocks}

In this section, we mention
the most concrete realization of integrability
of conformal blocks: in the case of the four external legs.
Namely its Fourier transform
\be
Z_N(\eta) =
\sum_{k=-\infty}^\infty \ Z_{k,N-k} \cdot e^{ik\eta}
\label{FT1}
\ee
satisfies the equation Painlev\'e VI, this was discovered in \cite{GIL1} as an interpretation of the old result of \cite{Jimbo}.
satisfies the Painlev\'e VI equation.
Here $e^{i\eta} = \frac{\mu_1}{\mu_2}$.
Summation over $k$ is actually over $N_1=\alpha-\alpha_1-\alpha_2$,
and, for integer
$N_1$ and $N$, it is automatically restricted to
the finite segment $0\leq k\leq N$ due to the factorials (Gamma-functions)
in the denominator of (\ref{ZB}).
Generically one can consider this is a sum over $\alpha$,
which parameterizes the internal dimension $\Delta=\alpha^2$:
\be
Z(\alpha,\eta) = \sum_{k=-\infty}^\infty   z(\alpha+k)\cdot e^{ik\eta}
\label{FT2}
\ee
and then it additionally depends on the non-integer part of $\alpha$.
Dependence on external dimensions $\alpha_1,\ldots,\alpha_4$ (including
not-obligatory-integer $N=-\sum_{i=1}^4\alpha_i$) is suppressed in this formula.

\bigskip

The Painlev\'e VI equation
is, in fact, a homogeneous equation in $Z$ of degree $4$,
but in these terms it is rather long.
Its condensed form used in \cite{GIL1} is in terms of
$\zeta(q) = q(q-1)\frac{\partial \log Z}{\partial q}$ and looks like
\be
\Big(q(q-1)\zeta''\Big)^2 = -2\det\left(
\begin{array}{ccc} 2\alpha_1^2 & q\zeta'-\zeta
& \zeta' + \alpha_1^2+\alpha_2^2+\alpha_3^2-\alpha_4^2 \\
q\zeta'-\zeta & 2\alpha_2^2 & (q-1)\zeta'-\zeta\\
\zeta' + \alpha_1^2+\alpha_2^2+\alpha_3^2-\alpha_4^2
& (q-1)\zeta'-\zeta & 2\alpha_3^2
\end{array}
\right)
\label{PVI}
\ee
where prime denotes a $q$-derivative (for conformal block $q$ is the position
of external leg, when the other two are located at $0$, $1$ and $\infty$,
or, in general a double ratio of the four positions).
Substitutions $\alpha_4^2=(\alpha_1+\alpha_2+\alpha_3)^2-\rho$ and
$\zeta =  \alpha_1\alpha_2 (q-1) + \alpha_2\alpha_3 q + \xi$ convert (\ref{PVI}) into
\be
\xi'\big(q\xi'-\xi\big)\Big((q-1)\xi'-\xi\Big)
+ \left(\frac{q(q-1)\xi''}{2}\right)^2
+2\alpha_2\rho\Big((\alpha_1-\alpha_3)(q\xi'-\xi) -(\alpha_1+\alpha_2)\xi'\Big)
-\alpha_2^2\rho^2+
\nn \\
+\underbrace{\big(\rho-(\alpha_1+\alpha_3)^2\big)}_{
\alpha_2^2+2\alpha_2(\alpha_1+\alpha_3)-\alpha_4^2}(q\xi'-\xi)^2
+\underbrace{(2\alpha_1^2+2\alpha_1\alpha_2+2\alpha_1\alpha_3-2\alpha_2\alpha_3-\rho)}_{
\alpha_1^2-\alpha_2^2-4\alpha_2\alpha_3-\alpha_3^2+\alpha_4^2}
\xi'(q\xi'-\xi)
-(\alpha_1+\alpha_2)^2(\xi')^2 = 0
\label{PVI'}
\ee
One can easily check that (\ref{detrep}) at $N=0$, i.e.
\be
Z_{N=0} = q^{2\alpha_1\alpha_2} (q-1)^{2\alpha_2\alpha_3}
\ \ \ \ \ \ \ \ \ {\rm with} \ \ \ \ \ \ \ \ N=-\sum_{i=1}^4\alpha_i=0
\ee
solves (\ref{PVI}): both sides of (\ref{PVI}) vanish in this case, while in (\ref{PVI'}) both
$\rho=0$ and $\xi=0$.
This trivial solution provides only a "perturbative" prefactor
in front of the conformal block at non-zero $N$.
However, one can make a computer check that the first terms of $q$-expansion (\ref{detrep}) at non-zero $N$, i.e. the Fourier transform
(\ref{FT1}) of the conformal block at arbitrary $N$, also satisfy (\ref{PVI}) (see also \cite{FW}).
Moreover, one can also check that (\ref{PVI}) is fulfilled iff the coefficients
in front of the poly-linear combinations of hypergeometric functions
are, indeed, unit, as implied by (\ref{preFT1}).
It is appealing to interpret {\bf (\ref{detrep}) as a kind of a non-linear
transform relating the Painlev\'e VI and the much simpler hypergeometric equation}.

The check that the Fourier transform (\ref{FT2}) satisfies (\ref{PVI}),
which was suggested in \cite{GIL1}, is less sophisticated:
one just looks for a solution of (\ref{PVI}) in the form (first proposed in \cite{Jimbo})
\be
Z=\sum_k e^{ik\eta}\cdot q^{(\alpha+k)^2-\alpha_1^2-\alpha_2^2}\cdot
 \sum_{i=0}F_i(\alpha+k)\cdot q^i
\ee
and realizes, term by term in $q$, that the ratios $F_i(\alpha)/F_0(\alpha) $
are nothing but the coefficients of expansion of the conformal block (\ref{CB}),
while $F_0(\alpha)$ is a product of the Barnes $G$-functions $\mathfrak{G}$.
We comment on this check  in a little bit simpler example of the Painlev\'e III equation in s.\ref{FTZP} below.
Two different {\it proofs} that (\ref{FT2}) satisfies (\ref{PVI})
were provided in \cite{ILT2} and \cite{BS1}. However, they are valid only for the case (\ref{cn1}), while the application/extension of these proofs to the mostly interesting case of integer $N_1$ and $N$ requires some care.

\bigskip

The Painlev\'e equation looks like a sophisticated non-linear equation of a rather strange form. However, it just a particular example of a set of Toda $\tau$-functions satisfying the usual bilinear Hirota relation \cite{Oka}
\be 
\tau_n\partial^2\tau_n-(\partial\tau_n)^2=\tau_{n+1}\tau_{n-1},\ \ \ \ \ \ \partial=q(1-q){\partial\over\partial q}
\ee 
and often possess determinant representations (see, for example, \cite{Yamada}). In fact, the Painlev\'e equation can be considered as a counterpart of the string equation, which picks up a distinguished subset of $\tau$-functions, and reflect the {\it super}-integrability of matrix models. We give a little more details about the interplay between the bilinear and Painlev\'e equations in discussion of a simpler Painlev\'e III example in sec.\ref{TI} below, which is associated with the pure gauge limit (PGL) of conformal blocks.
To describe the PGL at the level of the Painlev\'e equations, one makes a slightly different  substitution
$\zeta =  \alpha_1\alpha_2 (q-1) + \alpha_2\alpha_3 q +(\alpha_1+\alpha_2)^2- \xi$
which converts (\ref{PVI}) into
\be
\underline{\frac{1}{4}\Big({q}({q}-\underline{1}){\xi''}\Big)^2
- \xi'\big(q\xi'-\xi\big)\Big(({q}-\underline{1})\xi'-{\xi}\Big)}
-2\alpha_2(\alpha_1-\alpha_3)\rho (q\xi'-\xi)
-\alpha_2^2\rho^2+
\nn \\
+ {\big(\rho-(\alpha_1+\alpha_3)^2\big)} \Big(q\xi'-\xi+(\alpha_1+\alpha_2)^2\Big)^2
- {(2\alpha_2^2+ 2\alpha_1\alpha_2-2\alpha_1\alpha_+2\alpha_2\alpha_3+\rho)}
\xi'(q\xi'-\xi) + \nn\\
-2\alpha_2\rho (\alpha_1+\alpha_2)^2(\alpha_1-\alpha_3)
\Big)
- \underline{(\alpha_1^2-\alpha_2^2)(\alpha_3^2-\alpha_4^2)\xi'}
= 0
\ee
Underlined are the terms of the order $q^{-2}$ or $\alpha^4 q^{-1}$,
which survive in the pure gauge limit (PGL),
when $\alpha_{1,2,3,4}\longrightarrow \infty$ and
$q=\frac{t}{(\alpha_1^2-\alpha_2^2)(\alpha_3^2-\alpha_4^2)} \longrightarrow 0$
with finite $t$.

\section{PGL from Virasoro representation theory
\label{Shapo}}

We now switch to the theory in the pure gauge limit (PGL).
This is quite straightforward at the level of conformal blocks.

According to \cite{nconfAGT},  the 4-point block can be expressed through
inverse {\bf Shapovalov matrix} with $\Delta = \frac{n^2}{4}$:
\be
\boxed{
B_* = \sum_{n=0}^\infty \Lambda^{4n} \cdot Q^{-1}_\Delta\Big([1^n],[1^n]\Big)
}
\ee
(which allows one to treat it as a norm of the peculiar Gaiotto state \cite{Gaistate}).

For the sake of convenience, we list the first entries of Shapovalov matrix in the Appendix, where
boxed are the matrix elements contributing to the conformal block in the PGL. This gives the answer
\be
Z_*^{(1)} =  1 + \frac{2\Lambda^4}{n^2} + \frac{(2n^2+1)\,\Lambda^8}{n^2(n^2-1)^2}
+ \frac{2\,(2n^4-5n^2+12)\,\Lambda^{12}}{3\,n^2(n^2-1)^2(n^2-4)^2}
+ \frac{(4n^8-52n^6+243n^4-177n^2+324)\,\Lambda^{16}}{6\,n^4(n^2-1)^2(n^2-4)^2(n^2-9)^2}
+ O(\Lambda^{20})
\label{Z1expanfirsttermsSha}
\ee

\bigskip

For further applications, (\ref{Z1expanfirsttermsSha}) can be re-expanded as
\be
Z_*^{(1)} =  1 + \frac{2\Lambda^4}{n^2} + \left(\frac{1}{n^2} + \frac{3}{4(n+1)^2}
+ \frac{3}{4(n-1)^2} + \frac{5}{4(n+1)} - \frac{5}{4(n-1)}\right)\Lambda^8 +
\ee
{\footnotesize $$
\!\!\!\!\!\!\!\!\!\!\!\!\! \!\!\!\!\!\!
+ \left(\frac{1}{2n^2} + \frac{1}{6(n+1)^2}+\frac{1}{6(n-1)^2}
+ \frac{1}{36(n+2)^2} + \frac{1}{36(n-2)^2}
+\frac{17}{54(n+1)}-\frac{17}{54(n-1)}
+ \frac{7}{108(n+2)}-\frac{7}{108(n-2)}\right)\Lambda^{12} + O(\Lambda^{16})
$$}

\bigskip

\noindent
or
{\footnotesize
\be
Z_*^{(1)} =  1 + \frac{2\Lambda^4}{n^2} + \left(\frac{1}{n^2} -\frac{1}{n^2-1} +
\frac{3}{(n^2-1)^2}\right)\Lambda^8
+ \left(\frac{1}{2\,n^2} - \frac{8}{27\,(n^2-1)}+\frac{2}{3\,(n^2-1)^2}
- \frac{11}{54\,(n^2-4)} + \frac{4}{9\,(n^2-4)^2}\right)\Lambda^{12} +
\ee
$$
\!\!\!\!\!\!\!\!\!\!\!\!\!\!\!\!\!\!\!\!\!\!\!\!\!\!\!\!\!\!\!\!
+ \left(\frac{235}{32\cdot 81\, n^2}+\frac{1}{24\,n^4} - \frac{403}{256\cdot 27\,(n^2-1)}
+\frac{19}{64\cdot 3\,(n^2-1)^2}
- \frac{25}{32\cdot 27\,(n^2-4)} + \frac{1}{18\,(n^2-4)^2}
- \frac{71}{256\cdot 81\,(n^2-9)} + \frac{5}{64\cdot 9\, (n^2-9)^2}
\right)\Lambda^{16}  +
 O(\Lambda^{20})
$$
}
In this form it can be used to interpret the Fourier transform of conformal
block $n$ as s series
\be
{\rm FT(conf.block)} = \left(1+\sum_{k=0} \frac{\Lambda^{4k+4}\beta_k(\Lambda^4)}{{\bf d}-k^2} +
\sum_{k=0} \frac{\Lambda^{4k+4}\gamma_k(\Lambda^4)}{({\bf d}-k^2)^2}\right) \Theta
\ee
with ${\bf d} = d/d\log(\Lambda^4)$ and $\Theta$ depends on the
choice of the $U(1)$-prefactor in front of  (\ref{Z1expanfirsttermsSha}), but we will use a slightly different
method in s.\ref{FTZP} below.

\section{PGL via unitary models
\label{BGWMc}}

As further explained in \cite{ShaPGL}, this quantity
can be alternatively
expressed as a {\bf BGW matrix model}
\be\label{BGW}
Z_{BGW}(n|\Psi)= \frac{1}{{\rm Vol}_\beta(n)}\int [dU]_\beta e^{\beta(\Tr U^\dagger +
\Tr \Psi U)}
\ee
where $\beta$ refers to a $\beta$-deformation
of unitary integrals and volumes, which we do not need below,
because will actually deal only with the case of $\beta=1$.
The measure $[dU]$ is normalized to unity: $\int [dU] = 1$.

At $\beta = 1$,
\be\label{Z*}
Z_*^{(1)} = \int [dU] \int [dV] Z_{BGW}(m_+|U) Z_{BGW}(m_-|V)
\det\Big(1-\Lambda^4\ U^\dagger\otimes V^\dagger\Big)^2 =\nn\\
= \sum_{R,Q} \frac{d_R^2}{D_R(m_+)}\cdot\frac{d_Q^2}{D_Q(m_-)}\cdot
\sum_{X,Y}
(-\Lambda^4)^{|X|+|Y|}
\left(\int [dU] \chi_R[U] \chi_{_X}[U^\dagger]\chi_{_Y}[U^\dagger]\right)
\left(\int [dV] \chi_Q[V] \chi_{_{X^{tr}}}[V^\dagger]\chi_{_{Y^{tr}}}[V^\dagger]\right)
\ee
where $\chi_{_R}$ are the characters of the linear group (the Schur polynomials). Here we used that
\be
\det (1-\Lambda^4 \cdot U \otimes V)=\sum_R (-1)^{|R|}\chi_{_R}(U)\chi_{_R^{tr}}(V)
\ee
where $X^{tr}$ denotes the conjugated Young diagram, and the character expansion of (\ref{BGW}) valid at $\beta=1$:
\be
Z_{BGW}(n|U)=\sum_R{d_R^2\over D_R(n)}\chi_{_R}(U)
\ee

Since
\be
\chi_{_X}\chi_{_Y}=\sum_Z C_{_{Z,X,Y}} \cdot \chi_{_Z}
\ee
and
\be
\int_{n\times n} [dU]\ \chi_{_R}(U)\chi_{_Z}(U^\dagger) = \delta_{R,Z} \cdot \theta(n-l_R)
\ \ \ \ \ \ \ \ \ \ \ \ \ \ \ \theta(x)=
\left\{\begin{array}{ccc}
1 & {\rm for} & x\geq 0 \\  0 &{\rm for} & x<0
\end{array}
\right.
\ee
the matrix integral (\ref{Z*}) is actually equal to
\be
Z_*^{(1)} = \sum_{R,Q} (-\Lambda^4)^{|R|} \delta_{|R|,|Q|}\cdot K_{RQ}
\cdot \frac{d_R^2d_Q^2}{D_{_R}(m_+) D_{_Q}(m_-)}
\label{Z1expan}
\ee
with
\be
K_{RQ} = \sum_{X,Y} C_{_{R,X,Y}} C_{_{Q,X^{tr},Y^{tr}}}
\ee

Note that, since at $\beta=1$ the (analytically continued) "size" $m_-=-m_+=-n$
is negative, the expansion (\ref{Z1expan}) actually involves a transposed matrix $K$ defined as a reflection w.r.t. the vertical axis:
\be
\boxed{
Z_*^{(1)} \ \stackrel{\beta=1}{=}\ \sum_{R,Q} \Lambda^{4|R|} \delta_{|R|,|Q|}\cdot K_{RQ}
\cdot \frac{d_R^2\,d_Q^2}{D_{\!_R\!}(n) \,D_{\!_{Q^{tr}}\!}(n)}
= \sum_{R,Q} \Lambda^{4|R|} \delta_{|R|,|Q|}\cdot K_{RQ}^{tr}
\cdot \frac{d_R^2\,d_Q^2}{D_{\!_R\!}(n)\, D_{\!_Q\!}(n)}
}
\label{Z1expan_a}
\ee
The first examples of the matrices $K$, $K^{tr}$ can be found in the Appendix. Inserting them into
(\ref{Z1expan_a}) gives
\be
\boxed{
Z_*^{(1)} =  1 + \frac{2\Lambda^4}{n^2} + \frac{(2n^2+1)\,\Lambda^8}{n^2(n^2-1)^2}
+ \frac{2\,(2n^4-5n^2+12)\,\Lambda^{12}}{3\,n^2(n^2-1)^2(n^2-4)^2}
+ \frac{(4n^8-52n^6+243n^4-177n^2+324)\,\Lambda^{16}}{6\,n^4(n^2-1)^2(n^2-4)^2(n^2-9)^2}
+ O(\Lambda^{20}) =(\ref{Z1expanfirsttermsSha})
}\nn
\ee
At $\beta\neq 1$ ($c\neq 1$), the formulas are a little more involved,
and the corresponding characters
are the Jack rather than the Schur polynomials.

\bigskip

One can easily obtain explicit formulas for the matrices $K_{RQ}$ and $K_{RQ}^{tr}$ using the expansion of the characters (the Schur polynomials) in accordance with the Frobenius formula
\be
\chi_R (U)=\sum_\Delta {\psi_R(\Delta)\over z_\Delta}\prod_i\Tr U^{\delta_i}
=\sum_\Delta d_R\varphi_R(\Delta) \prod_i\Tr U^{\delta_i}
\ee
where $\Delta$ is the Young diagram with $l(\Delta)$ lines with lengths $\delta_1\ge\delta_2\ge\ldots\delta_{l(\Delta)}\ge 0$ so that $|\Delta|=|R|$, $\psi_R(\Delta)$ is the character of the symmetric group $S_{|R|}$, and $z_\Delta$ is the standard symmetric factor of the Young diagram $\Delta$ (order of automorphism) \cite{Fulton}. One will also need the orthogonality relations
\be
\sum_R\psi_R(\Delta_1)\psi_R(\Delta_2)=z_{\Delta_1}\delta_{\Delta_1,\Delta_2}\\
\sum_R\psi_R(\Delta_1)\psi_{R^{tr}}(\Delta_2)=(-1)^{l(\Delta_1)+|\Delta_1|}z_{\Delta_1}\delta_{\Delta_1,\Delta_2}
\ee
in order to obtain
\be
K_{RQ}=(-1)^{|R|}\sum_{\Delta_1,\Delta_2:\ |\Delta_1|+|\Delta_2|=|R|}(-1)^{l(\Delta_1+\Delta_2)}{\psi_R(\Delta_1+\Delta_2)\psi_Q(\Delta_1+\Delta_2)\over z_{\Delta_1}z_{\Delta_2}}
 \nn \\
K_{RQ}^{tr}=
\sum_{\Delta_1,\Delta_2:\ |\Delta_1|+|\Delta_2|=|R|}
{\psi_R(\Delta_1+\Delta_2)\psi_Q(\Delta_1+\Delta_2)\over z_{\Delta_1}z_{\Delta_2}}
\ee
Here the sum of two Young diagrams $\Delta_1+\Delta_2$ is defined to be an ordered set of union of lines of the two diagrams and summation includes $\Delta_1=\emptyset$ and $\Delta_2=\emptyset$.

A more interesting question is what is a non-technical reason for Shapovalov
and Littlewood-Richardson  formalisms to give the same answers,
it remains beyond the scope of the present paper.

\section{Fourier transform and Painlev\'e III
\label{FTZP}}

Since integrability behind the conformal blocks (\ref{detrep}) gets explicit
only after the Fourier transform in the internal $\alpha$-parameter,
one can expect the same to happen in the pure gauge limit.
This expectation is indeed true as observed recently in \cite{GIL3,ILT3,BS1,GL}.
In what follows, we describe our understanding of this story.

That is, the Fourier transform of (\ref{Z1expanfirsttermsSha}),
the PGL of conformal block satisfies
the equation Painlev\'e III,
which can be written in many different forms \cite{BS3}.  In the PGL, (\ref{PVI'}) turns into
\be\label{Peq}
\frac{1}{4}(t\ddot\zeta )^2+\dot\zeta^2(t\dot\zeta-\zeta)= -\dot\zeta
\ee
where dot denotes the derivative w.r.t. $t=\Lambda^4$ and
$\zeta = t\frac{d}{dt}\log Z$.
In fact, this is a quartic homogeneous equation in $Z$:
$$
 t^4 (Z^2\dddot Z^2 +4Z\ddot Z^3-6Z\dot Z\ddot Z\dddot Z-3\dot Z^2\ddot Z^2
+4\dot Z^3\dddot Z)
+4t^3(Z^2\ddot Z\dddot Z -Z\dot Z\ddot Z^2 -Z\dot Z^2\dddot Z +\dot Z^3\ddot Z)
+4t^2(Z^2\ddot Z^2 - Z\dot Z^2\ddot Z)= $$
\be
= 4t(Z^2\dot Z^2 - Z^3\ddot Z) - 4Z^3\dot Z
\label{Zeq}
\ee
(note that the two sides are also homogeneous in $t$, but of different degrees,
$-2$ at the l.h.s. and $-1$ at the r.h.s.).

Following \cite{Jimbo,GIL2,ILT3,BS1,GL},
we look for its solution in the form of a Fourier transform (\ref{FT2}) of some
series $\sum_i F_i(a)\cdot t^i$:
\be\label{FZ}
Z=\sum_{k\in\mathbb{Z}} t^{(a+k)^2}\cdot \left(\sum_{i=0}F_i(a+k)\cdot t^i\right)
\ee
Because of the presence of $a$, which does not need to be integer, in the exponential,
this is actually a double series in integer powers of two independent
parameters $t$ and $t^{2a}$.
Thus vanishing should be all the coefficients of this {\it double} expansion,
i.e. coefficients in front of any $t^{4a^2\pm 2k_1a +k_2}$ with
$k_1,k_2\in\mathbb{Z}_{\ge 0}$.
This imposes an enormously big set of constraints on the functions $F_i(a)$,
but it has a solution.
To illustrate how this works,
consider, for example, the coefficient in front of $t^{4a^2+2a}$ to see that
\be\label{c1}
F_1(a)={1\over 2a^2}F_0(a)
\ee
It simultaneously cancels the coefficient of $t^{4a^2+4a+1}$.
Similarly, looking at $t^{4a^2}$, one obtains that
\be\label{c2}
F_0(a+1)F_0(a-1)=\Big[{1\over 4a^2(4a^2-1)}\Big]^2F_0(a)^2
\ee
The same condition cancels the coefficient in front of $t^{4a^2+1}$. The next degrees already give a condition for $F_2(a)$: vanishing the coefficient in front of $t^{4a^2+2a+1}$ gives
\be\label{c3}
F_2(a)={(8a^2+1)\over 4a^2(4a^2-1)^2}F_0(a)
\ee
which simultaneously guarantees cancelling of $t^{4a^2+2}$ and $t^{4a^2+4a+2}$. Further, one can determine from the coefficient of $t^{4a^2+2a+2}$ that
\be\label{c4}
F_3(a)={8a^4-5a^2+3\over 24 a^2(a-1)^2(2a+1)^2(2a-1)^2(a+1)^2}F_0(a)
\ee
Note that for these calculations it was sufficient to keep only Fourier modes
with  $k=0,\pm 1,\pm 2$ in (\ref{FZ}).
Note also that (\ref{c2}) implies that
\be
F_0(a)={1\over \mathfrak{G}(1+2a)\mathfrak{G}(1-2a)}
\ee

As soon as the polynomial in front of $t^{4a^2+2ka}$ vanishes at each $k$ separately,
one obtains that the equation (\ref{Zeq}) is satisfied by a more general function
\be\label{FZf}
Z=\sum_{k\in\mathbb{Z}} t^{(a+k)^2}\sum_{i=0}F_i(a+k)\cdot t^i\cdot e^{ik\eta}
\ee

It remains to note that the coefficients of the expansion (\ref{c1}), (\ref{c3}), (\ref{c4})
coincide with those of $Z_*^{(1)}$ in (\ref{Z1expanfirsttermsSha}),
i.e. finally the solution to the equation (\ref{Zeq}) can be written as
\be
Z=\sum_{k\in\mathbb{Z}} \mathfrak{Z}(a+k)e^{ik\eta}\,,\ \ \ \ \ \ \ \ \ \ \ \ \
\mathfrak{Z}={t^{a^2}Z_*^{(1)}(n=2a)\over \mathfrak{G}(1+2a)\mathfrak{G}(1-2a)}
\ee
or the Fourier transform of {\bf the PGL conformal block satisfies the Painlev\'e III equation}, as claimed in \cite{GIL2,ILT3,BS1,GL}.

\section{Relation to Toda integrability\label{TI}}

Our last task in the present paper is
to explain what has Painlev\'e III to do with the ordinary KP/Toda integrability,
typical for the eigenvalue matrix models.

For this we note that the homogeneity of the equation (\ref{Zeq})
makes it much similar to the Hirota equation.
However, the Hirota equation is bilinear, while (\ref{Zeq}) is quartic.
Hence, one may expect that there is a bilinear B\"acklund transformation
to another function $Z_1$, in an analogy with the mKdV case, when
the standard infinite set of Hirota equations for the Toda $\tau$-function
\be\label{Tc}
\tau_n\partial^2\tau_n-(\partial\tau_n)^2=\tau_{n+1}\tau_{n-1}
\ee
reduces to a pair of equations
\be
\tau_0\partial^2\tau_0-(\partial\tau_0)^2=\tau_1^2\nn\\
\tau_1\partial^2\tau_1-(\partial\tau_1)^2=\tau_0^2
\label{doubleq}
\ee
provided by the reduction
\be
\tau_{n+2}=\tau_n
\ee
Indeed,
it turns out  that the equation (\ref{Zeq}) can be rewritten in a much similar form of two equations \cite{BS3}
\be
Z\partial^2Z-(\partial Z)^2=tZ_1^2\nn\\
Z_1\partial^2Z_1-(\partial Z_1)^2=Z^2
\label{doubleqZ}
\ee
with a B\"acklund transformed function $Z_1$.
The derivative here is taken w.r.t. $\log t$: $\partial={\partial\over\partial\log t}$.
Because of this one can easily multiply $Z$ or $Z_1$ by powers of $t$,
and change the pair of coefficients at the r.h.s. from $(t,1)$ to, say $(t^{1/2},t^{1/2})$.
Moreover, one can obtain such $Z$ from
a quasi-periodic solution\footnote{
It can be also obtained
as an automodel solution of the sine-Gordon or 2-periodic two-dimensional Toda equation \cite{BS3}.
}
to the Toda chain (\ref{Tc}) with the periodicity condition $\tau_{n+2}=\sqrt{t}\cdot \tau_n$, then $Z=\tau_1$ and  $Z_1=t^{1/4}\cdot\tau_0$.
To prove the equivalence of (\ref{doubleq}) to (\ref{Peq}) one needs to express
$Z_1$ through $Z$ from the first equation:
\be
Z_1^2={1\over t}\,Z^2\cdot\partial^2\log Z=Z^2\cdot{\partial \zeta\over\partial t}
\label{Z1}
\ee
and then substitute into the second equation.
Emerging equation differs from (\ref{Zeq}), in particular,
it contains several terms with the fourth derivative of $Z$.
They can be easily eliminated, because the
derivative of (\ref{Peq}) factorizes:
\be
\left(t^2\dddot\zeta + t\ddot\zeta + 6t\dot\zeta^2-4\zeta\dot\zeta+2\right)
\cdot\ddot\zeta =0
\ee
This allows one to express $\dddot\zeta$ and thus the forth derivative of $Z$
through lower derivatives and check that (\ref{doubleqZ}) is indeed
equivalent to (\ref{Zeq}).

For $Z$ of the form (\ref{FZf}), the B\"acklund transform (\ref{Z1}) implies:
\be
Z_1=t^{a^2}\left[{C_0F_0(a)\over a}\left(1+{2(4a^2+1)\over (4a^2-1)^2}t+...\right)+aC_1F_0(a+1)t^{2a}\left((2a+1)^2+{2(4a^2-8a+1)\over (2a-1)^2}t+...\right)+\right.\nn\\
+aC_1F_0(a-1)t^{-2a}\left((2a-1)^2+{2(4a^2+8a+1)\over (2a+1)^2}t+...\right)+\nn\\
+a^3(2a+1)^2C_2t^{4a}{F_0(a+1)^2\over F_0(a)}\left((2a+1)^2+{128(4a^2-16a+1)\over (2a-1)^2}t+...\right)+\nn\\
\left.+a^3(2a-1)^2C_2t^{-4a}{F_0(a-1)^2\over F_0(a)}\left((2a-1)^2+{128(4a^2+16a+1)\over (2a+1)^2}t+...\right)+...\right]
\ee

Here the constants $C_i$ are expressed through the root $x$ of the equation $49x^4-308x^3+580x^2-128x+2=0$:
\be
C_0&=&\sqrt{x}\ {29400x^3-185927x^2+355476x-90512\over 17084}\nn\\
C_1&=&\sqrt{x}\ {5635x^3-37380x^2+77102x-28168\over 4271}\nn\\ \nn \\
C_2&=&4\sqrt{x} \nn \\ \nn \\
\ldots
\ee

\section{Conclusion}

In this paper, we reminded a piece of the old theory from \cite{UFN3}
and \cite{AGTmamo,MMSha} and once again emphasized
the importance of matrix model techniques for modern studies of the AGT relations
and  other subjects in representation theory.
That is, we explained that because conformal blocks are described by
non-trivial (Dijkgraaf-Vafa \cite{DV}) phases of conformal matrix models,
their integrability
can be seen only after a Fourier transform in (square roots of)
the intermediate dimensions.
We also stressed that the Painlev\'e equations, which these Fourier transforms
were discovered to satisfy in \cite{GIL1}-\cite{GL},
can in fact be naturally embedded (which is, in no way, a surprise) into the KP/Toda context
usual for matrix models.
Moreover, we argued that not only a rather formal Fourier transform of \cite{GIL1},
but also the very explicit  one, necessarily emerging from the matrix model representation of the conformal block and expressed as a poly-linear combination
of hypergeometric functions, satisfies the same Painlev\'e VI equation.
This reveals a connection between the Painlev\'e and hypergeometric equations
which deserves separate investigation.

Further, we considered the pure gauge limit addressed also in \cite{GIL2,ILT3,BS1,GL}.
What we especially like about the recent \cite{GL} is that it addresses
the subject actually related to the Brezin-Gross-Witten (BGW) model,
which does not attract attention it deserves, especially, among other matrix models.
Hopefully, \cite{GL} together with our comments in the present paper
would help to change this attitude.
From the point of view of Painlev\'e theory, this case is even simpler,
because emerging is the Painlev\'e III equation rather than the usual Painlev\'e VI one.
For all these reasons, we thoroughly considered the PGL example in the present paper.
For further developments in the theory of BGW matrix models, see \cite{AMM}. For an alternative matrix model description, see \cite{Grassi}.

Other obvious next steps in the study include:

\begin{itemize}
\item Generalization from 4-point to arbitrary conformal blocks, which is absolutely immediate in terms of determinant representations and integrability, however, counterparts of the Painlev\'e equations still need to be found (cf. \cite{GL1});
\item Generalization to $c\neq 1$: as mentioned at the end of s.\ref{detrepDF},
determinant representation is lost,
but the idea of Fourier transform survives and it remains to work out
a language adequate for application to $\beta$-ensembles;
\item Further generalization to the balanced networks and other DIM-related
models of \cite{DIMmods}, which is related to the $5d$ generalizations of the AGT correspondence \cite{5d} and $q$-Painlev\'e \cite{Y,BS2,JNS};
\item Generalization to elliptic/toric conformal blocks of \cite{MMShaell,Y2}.
\end{itemize}

\section*{Acknowledgements}

We acknowledge the stimulating atmosphere of the VII Workshop on Geometric Correspondences of Gauge Theories organized by Giulio Bonelli and Alessandro Tanzini at SISSA, and useful comments by its numerous participants, especially by Misha Bershtein, Alba Grassi, Yasuhiko Yamada and Yegor Zenkevich.

This work was performed at the Institute for Information Transmission Problems with the
financial support of the Russian Science Foundation (Grant No.14-50-00150).

\section*{Appendix}

In this Appendix, we list the Shapovalov matrices and the $K$-, $K^{tr}$-matrices at the first 4 levels. These formulas are necessary for reproducing the first terms of expansion of the PGL conformal block.

\paragraph{The Shapovalov matrices and their inverse.}

\be
\text{level 1:}  & \!\!\!\!\!\!\!\!\!\!\!\!\!\!\!\!\!\!\!\!\!\!\!\!\!\!\!\!\!\!\!\!\!
\!\!\!\!\!\!\!\!\!\!\!\!\!\!\!\!\!\!\!\!\!\!\!\!\!\!\!\!\!\!\!\!\!
\!\!\!\!\!\!\!\!\!\!\!\!\!\!\!\!\!\!\!\!\!\!\!\!\!\!\!\!\!\!\!\!\!
Q = 2\Delta
& \!\!\!\!\!\!\!\!\!\!\!\!\!\!\!\!\!\!\!\!\!\!\!\!\!\!\!\!\!\!\!\!\!
\!\!\!\!\!\!\!\!\!\!\!\!\!\!\!\!\!\!\!\!\!\!\!\!\!\!\!\!\!\!\!\!\!
\!\!\!\!\!\!\!\!\!\!\!\!\!\!\!\!\!\!\!\!\!\!\!\!\!\!\!\!\!\!\!\!\!
\!\!\!\!\!\!\!\!\!\!\!\!\!\!\!\!\!\!\!\!\!\!\!\!\!\!\!\!\!\!\!\!\!
Q^{-1} = \frac{1}{2\Delta} = \boxed{\frac{2}{n^2}} \nn\\ \nn \\
\text{level 2:} & \!\!\!\!\!\!\!\!\!\!\!\!\!\!\!\!\!\!\!\!\!\!\!\!\!\!\!\!\!\!\!\!\!
\!\!\!\!\!\!\!\!\!\!\!\!\!\!\!\!\!\!\!\!\!\!\!\!\!\!\!\!\!\!\!\!\!
\!\!\!\!\!\!\!\!\!\!\!\!\!\!\!\!\!\!\!\!\!\!\!\!\!\!\!\!\!\!\!\!\!
\!\!\!\!\!\!\!\!\!\!\!\!\!\!\!\!\!\!\!\!\!\!\!\!\!\!\!\!\!\!\!\!\!
Q =\frac{1}{2} \left(\begin{array}{c|cc} & [2] & [1,1] \\ \hline & \\
\phantom. [2] & 2n^2+1 & 3n^2 \\ \phantom. [1,1] & 3n^2 & n^2(n^2+2) \end{array}\right)
&  \!\!\!\!\!\!\!\!\!\!\!\!\!\!\!\!\!\!\!\!\!\!\!\!\!\!\!\!\!\!\!\!\!
\!\!\!\!\!\!\!\!\!\!\!\!\!\!\!\!\!\!\!\!\!\!\!\!\!\!\!\!\!\!\!\!\!
\!\!\!\!\!\!\!\!\!\!\!\!\!\!\!\!\!\!\!\!\!\!\!\!\!\!\!\!\!\!\!\!\!
\!\!\!\!\!\!\!\!\!\!\!\!\!\!\!\!\!\!\!\!\!\!\!\!\!\!\!\!\!\!\!\!\!
\!\!\!\!\!\!\!\!\!\!\!\!\!\!
Q^{-1}= \frac{1}{n^2(n^2-1)}\left(\begin{array}{c|cc} & [2] & [1,1] \\ \hline & \\
\phantom. [2] & n^2(n^2+2) & -3n^2 \\ \phantom. [1,1] & -3n^2 &\boxed{2n^2+1}
\end{array}\right) \nn \\ \nn \\
{\rm level \ 3:} & \!\!\!\!\!\!\!\!\!\!\!\!\!\!\!\!\!\!\!\!\!\!\!\!\!\!\!\!\!\!\!\!\!
Q= \frac{1}{4} \left(\begin{array}{c|ccc} &[3] & [2,1] & [1,1,1] \\ \hline & \\
\phantom. [3] & 2(3n^2+4) & 8(2n^2+1) & 24n^2\\
\phantom. [2,1] & 8(2n^2+1) & 2n^4+35n^2+8 & 9n^2(n^2+4) \\
\phantom. [1,1,1] & 24n^2 &9n^2(n^2+4) & 3n^2(n^2+2)(n^2+4)\end{array}\right)\nn\\ \nn\\
& \!\!\!\!\!\!\!\!\!\!\! Q^{-1}= \frac{1}{3n^2(n^2-1)^2(n^2-4)^2}
\left(\begin{array}{c|ccc} &[3] & [2,1] & [1,1,1] \\ \hline &\\
\phantom. [3] & 2(n^2-1)^2(n^2+4)(n^2+8) & -16(n^2-1)^2(n^2+4)& 32(n^2-1)^2 \\
\phantom. [2,1] & -16(n^2-1)^2(n^2+4) & 6n^6+44n^4-96n^2+64 & -18n^4+32n^2-32 \\
\phantom. [1,1,1] & 32(n^2-1)^2 &-18n^4+32n^2-32 & \boxed{4n^4-10n^2+24}
\end{array}\right) \nn \\
\ldots \nn
\ee
Boxed are the matrix elements, contributing to conformal block in the PGL.

\paragraph{Matrices $K$ and $K^{tr}$.}

\be
{\rm level \ 1:} & K_{[1],[1]} = 2 \nn \\
{\rm level \ 2:} &
K= \left(\begin{array}{c|cc} & [2] & [1,1] \\ \hline
\phantom. [2] & 1 & 3 \\ \phantom. [1,1] & 3 &1 \end{array}\right)
& K^{tr}= \left(\begin{array}{c|cc} & [2] & [1,1] \\ \hline
\phantom. [2] & 3 & 1 \\ \phantom. [1,1] & 1 &3  \end{array}\right) \nn \\
{\rm level \ 3:} &
K= \left(\begin{array}{c|ccc} &[3] & [2,1] & [1,1,1] \\ \hline
\phantom. [3] & 0 & 2 & 4\\
\phantom. [2,1] & 2 & 6 & 2 \\ \phantom. [1,1,1] & 4 &2 & 0\end{array}\right)
& K^{tr}= \left(\begin{array}{c|ccc} &[3] & [2,1] & [1,1,1] \\ \hline
\phantom. [3] & 4 & 2& 0 \\
\phantom. [2,1] & 2 & 6 & 2 \\ \phantom. [1,1,1] & 0 &2 & 4 \end{array}\right)
\nn
\ee
\be
\!\!\!\!\!\!\!\!\!\!\!\!\!\!\! \!\!\!\!\!\!\!\!\!\!\!\!\!\!\! \!\!\!\!\!\!\!\!\!\!\!\!\!\!\!
\!\!\!\!\!\!\!\!\!\!\!\!\!\!\! \!\!\!\!\!\!\!\!\!\!\!\!\!\!\!
{\rm level \ 4:}
&K= \left(\begin{array}{c|ccccc} &[4] & [3,1] & [2,2]& [2,1,1] &[1,1,1,1] \\ \hline
\phantom. [4] & 0 &0 & 1 & 3 & 5\\
\phantom. [3,1] & 0 & 4 & 3 & 9 & 3\\
\phantom. [2,2] & 1 & 3& 6 & 3 & 1 \\
\phantom. [2,1,1] & 3 & 9 & 3 & 4 & 0 \\
\phantom. [1,1,1,1] & 5 &3 & 1 & 0 & 0 \end{array}\right)
& \  K^{tr}= \left(\begin{array}{c|ccccc} &[4] & [3,1] & [2,2]& [2,1,1] &[1,1,1,1] \\ \hline
\phantom. [4] & 5 &3 & 1 & 0 & 0\\
\phantom. [3,1] & 3 & 9 & 3 & 4 & 0 \\
\phantom. [2,2] & 1 & 3& 6 & 3 & 1 \\
\phantom. [2,1,1] & 0 & 4 & 3 & 9 & 3 \\
\phantom. [1,1,1,1] & 0 &0 & 1 & 3 & 5\end{array}\right)
\nn \\ \nn \\
& \ldots \nn
\ee

\end{document}